\documentclass[english]{DDCLSconf}
\usepackage[comma,numbers,square,sort&compress]{natbib}
\usepackage{epstopdf}
\usepackage{amsmath}
\usepackage{graphicx}
\usepackage{amsfonts}
\usepackage{hyperref}
\begin{document}

\title{Beyond Nash Equilibrium: Achieving Bayesian Perfect Equilibrium with Belief Update Fictitious Play}

\author{
  Qi Ju\aref{hust,exp},
  Zhemei Fang\aref{hust,exp},
  Yunfeng Luo\aref{hust,exp}
}


\affiliation[hust]{School of Artificial Intelligence and Automation,
        Huazhong University of Science and Technology, Wuhan 430074, P.~R.~China
        \email{juqi@hust.edu.cn}
}
\affiliation[exp]{National Key Laboratory of Science and Technology on Multispectral Information Processing, Wuhan 430074, P.~R.~China
        \email{juqi@hust.edu.cn}}
\maketitle

\begin{abstract}
  In the domain of machine learning and game theory,
  the quest for Nash Equilibrium (NE) in extensive-form games with incomplete information is challenging yet crucial for enhancing AI's decision-making support under varied scenarios.
  Traditional Counterfactual Regret Minimization (CFR) techniques excel in navigating towards NE,
  focusing on scenarios where opponents deploy optimal strategies.
  However,
  the essence of machine learning in strategic game play extends beyond reacting to optimal moves;
  it encompasses aiding human decision-making in all circumstances.
  This includes not only crafting responses to optimal strategies but also recovering from suboptimal decisions and capitalizing on opponents' errors.
  Herein lies the significance of transitioning from NE to Bayesian Perfect Equilibrium (BPE),
  which accounts for every possible condition,
  including the irrationality of opponents.

  To bridge this gap,
  we propose Belief Update Fictitious Play (BUFP),
  which innovatively blends fictitious play with belief to target BPE,
  a more comprehensive equilibrium concept than NE.
  Specifically,
  through adjusting iteration stepsizes,
  BUFP allows for strategic convergence to both NE and BPE.
  For instance,
  in our experiments, BUFP(EF) leverages the stepsize of Extensive Form Fictitious Play (EFFP) to achieve BPE,
  outperforming traditional CFR by securing a 48.53\% increase in benefits in scenarios characterized by dominated strategies.
\end{abstract}


  \footnotetext{This work is supported by National Natural Science Foundation of China(62103158). And this paper has been accepted by DDCLS 2024.}

  \section{Introduction}\label{sec:introduction}

Using artificial intelligence to solve game theory problems is an important branch in the field of machine learning.
In the incomplete information extensive-form game,
the current mainstream method is the counterfactual regret minimization (CFR) algorithm\cite{zinkevich2007regret, brown2019deep, lanctot2009monte, tammelin2014solving} based on the improvement of regret matching (RM) algorithm\cite{hart2000simple}.
However,
the potential of Fictitious Play (FP)\cite{brown1951iterative,fudenberg1998theory},
a classical game solving algorithm in normal-form games,
has not been fully explored.

CFR, while adept at finding Nash Equilibrium (NE),
does not extend to other types of equilibrium with the same ease.
In contrast,
FP offers a more versatile approach by enabling convergence to various equilibrium points through the adjustment of stepsizes.
However,
the difficulty of determining the best response strategy during iterative updates has limited FP's application in complex, large-scale games.
Motivated by the Bayesian Action Decoder (BAD)\cite{foerster2019bayesian} and Full-Width Extensive Form Fictitious Play (XFP)\cite{heinrich2015fictitious} algorithms,
we propose an innovative adaptation of FP — Belief Update Fictitious Play (BUFP).

BUFP distinguishes itself by meticulously tracking both the likelihood of reaching each information set and the player’s beliefs within those sets — aspects not fully accounted for by CFR.
Drawing inspiration from BAD,
BUFP simplifies solving the best response strategy across the game tree by focusing on individual subgame trees,
thereby significantly reducing computational complexity.
Furthermore, our analysis confirms that BUFP operates as a Generalized Weakened Fictitious Play (GWFP) process,
showcasing equivalent convergence properties to traditional FP algorithms.

A notable innovation of BUFP is its capacity to align with Bayesian Perfect Equilibrium (BPE) when employing a stepsize analogous to that of Extensive-Form Fictitious Play (EFFP)~\cite{hendon1996fictitious}.
This attribute ensures that BUFP(EF) optimizes returns, surpassing those of NE,
particularly when opponents deviate from optimal strategies.
This feature is empirically validated in our experiments within the 5-Leduc poker framework,
where BUFP(EF) outperforms CFR by approximately 48.53\% in scenarios where opponents err.

In the subsequent section,
we delve into a comprehensive review of pertinent literature in Section 2.
Section 3 is dedicated to outlining the foundational concepts critical to this study.
Section 4 provides a detailed examination of the BUFP methodology.
Section 5 presents a thorough analysis of experiments conducted.
Finally, the concluding section underscores the significant contributions of BUFP and proposes directions for further research.
Our code is available in \href{https://github.com/Zealoter/BUFP}{github} .

  \section{Related Work}\label{sec:related-work}

Algorithms designed for tackling extensive-form games fall into two principal categories:
conventional tabular methods and those that harness the power of deep learning.
Deep learning-based strategies have demonstrated remarkable success in complex gaming environments,
such as Go~\cite{silver2017mastering}, and StarCraft II~\cite{vinyals2019grandmaster}.
These cutting-edge techniques are grounded in the core principles of tabular algorithms but extend their capabilities through deep learning to navigate and manage the extensive strategic landscapes inherent in these games.

Within the field of tabular equilibrium algorithms
CFR~\cite{zinkevich2007regret} has emerged as a pivotal technique.
Evolving from RM~\cite{hart2000simple},
CFR disaggregates total regret across individual information sets,
optimizing each to minimize collective regret.
Enhancements to CFR have traditionally centered on enhancing its convergence rate.
The development of CFR+~\cite{tammelin2014solving}, for instance, marked a significant milestone by facilitating victories over professional human players in Texas Hold'em~\cite{bowling2015heads,brown2017safe, brown2019superhuman}.
Further advancements include Predictive CFR~\cite{farina2021faster},
which accelerates convergence by anticipating the payoffs of future actions,
and Dynamic Weighting-Pure CFR~\cite{ju2024integrating}, which further optimizes convergence speed through the dynamic adjustment of iteration weights.

In addition to the CFR algorithm,
Fictitious Play (FP)\cite{brown1951iterative} is another foundational tabular method to solved equilibrium.
However,
FP's dependency on a game's normal-formulation and the precise calculation of the BR strategy has historically limited its application.
Fudenberg\cite{fudenberg1998theory} comprehensive review of FP research delineates the algorithm's structure and evaluates its efficacy across various game types.
Hendon~\cite{hendon1996fictitious} adapted FP for extensive-form games.
David S. Leslie~\cite{leslie2006generalised} introduced the Generalized Weakened Fictitious Play (GWFP) process,
rigorously demonstrating its convergence to Nash equilibrium under certain conditions of perturbations and errors, akin to traditional FP.
Building upon EFFP and GWFP, Heinrich et al.
developed the XFP algorithm~\cite{heinrich2015fictitious},
enhancing FP's convergence rate to a game's weak solution in extended formats.

The cooperative board game Hanabi has emerged as a compelling subject of study~\cite{bard2020hanabi},
with the BAD algorithm~\cite{foerster2019bayesian} and its refinements achieving noteworthy success.
BAD's innovation lies in its dual recording of strategies and opponent beliefs at each information set,
utilizing this combined insight to refine the average strategy's update mechanism.
Given Hanabi's alignment with game theory's concept of strict dominant strategies,
it's conceivable that FP is well-equipped to tackle such challenges~\cite{nachbar1990evolutionary}.
The integration of Bayesian beliefs in Hanabi has demonstrated enhanced convergence to equilibrium,
suggesting a potentially broad applicability of belief recording in FP solvers.
If belief updates in Hanabi facilitate closer approximations to epsilon equilibrium,
it's plausible that incorporating belief updates in FP could similarly enhance outcome fidelity.

  \section{Preliminaries}\label{sec:preliminaries}

\subsection{Normal-Form and Extensive-Form Games}\label{subsec:normal-form-and-extensive-form-games}
Extensive-form games model sequential interactions in multi-agent environments,
consists of the following elements:
$\mathcal{N} = \{1, \dots, n\} $ represents the set of players.
All nodes in a finite game tree represent possible states $s\in\mathcal{S}$ in a game.
The leaf node $z\in \mathcal{Z}$ of the game tree is also called the terminal state.
For each state $s \in \mathcal{S}$,
its successor edge defines a set of actions $\mathcal{A}(s)$ that the player or dealer can take in the state $s$.
The player function $P : \mathcal{S} \rightarrow \mathcal{N} \cup \{c\}$ determines who acts in a given state,
where $c$ represents the chance.

In a game with imperfect information,
agent $i$ could only be aware of the several $s$ it is in and unable to point out which specific state it is in.
For each player $i$ there is a corresponding set of information states $\mathcal{U}^i$ and an information function
$ I^i :\mathcal{S} \rightarrow\mathcal{U}^i $ that determines which states are indistinguishable for the player
by mapping them on the same information state$ u \in \mathcal{U}^i $,
$\mathcal I(u)=\{s|s\in u\}$ represents the set of all possible nodes on the information set $u$.
Finally,
define the payoff function,
$R:\mathcal{Z}\rightarrow\mathbb{R}^n$ maps the terminal state to a vector,
its components correspond to each player's payoff.

A player's behavioral strategy $\pi^i(u)$ is a probability distribution on the action set $\mathcal{A}(u)$,
$\Delta^i_b$ is the set of all behavioral strategies of player $i$.
The strategy profile $\pi = (\pi^1, \dots, \pi^n)$ is the strategy profile of all players.
$\pi^{-i}$ refers to all policies in $\pi$ except $\pi^i$.
If all players follow the strategy profile $\pi$,
the game-based payoff function $R$ can get the expected payoff $R^i(\pi)$ for player $i$.
The BR strategy of player $i$ to his opponent is:

\begin{equation}
    \label{eq:p201}
    BR^{i}\left(\pi^{-i}\right)\in \arg \max _{\pi^{i} \in \Delta_{b}^{i}} R^{i}\left(\pi^{i}, \pi^{-i}\right)
\end{equation}

Note that there can be more than one optimal response,
and $BR(\pi)=\times_{i=1}^n BR^i(\pi^{-i})$ represents the strategy profile of all the BR strategies.
If an error $\epsilon > 0$ is allowed for the BR strategy, then:

\begin{equation}
    \label{eq:p202}
    \begin{split}
        BR_{\epsilon}^{i}&\left(\pi^{-i}\right)=\\
        &\left\{\pi^{i} \in \Delta_{b}^{i}: R^{i}\left(\pi^{i}, \pi^{-i}\right) \geq\right.\left.R^{i}\left(BR^{i}\left(\pi^{-i}\right), \pi^{-i}\right)-\epsilon\right\}
    \end{split}
\end{equation}

Every extensive-form game produces an equivalent normal-form game.
If only one of $\forall u\in \mathcal{U},\pi^i(u)$ is $1$ and the rest is $0$,
it is called a pure strategy marked as $\Pi^i$.

The cartesian product of all pure strategies constitutes the action set of player $i$ in a normal-form game:
$\Delta^i_p=\times_{u\in \mathcal{U}}\{a^s:a^s\in \mathcal{A}(u) ,P(u)=i\}$.
Define the mixed strategy $\Phi^i$ to represent the probability distribution of player $i$ in the action set $\Delta^i_p$ of the normal-form game.
Strategic game mixed strategy profile $\Phi =\{\Phi^1,\Phi^2,\cdots,\Phi^n\}$ is the mixed strategy profile of all players.
Finally,
$R^{i}(\Phi)$ determines the expected payoff of player $i$ for a given mixed strategy profile $\Phi$.

Throughout the paper,
we assume that the game has perfect recall,
and we use small Greek letters for extensive-form behavioral strategies and large Greek letters for pure and mixed strategies of normal-form game.

\subsection{Dominated Strategy}\label{subsec:dominated-strategy}
In game theory,
strategic dominance occurs when one action (or strategy) is better than another action (or strategy) for one player,
regardless of the opponents' actions.
Formally,
For player $i$,
if there is a pure strategy $a^{i*}$ and a strategy $\pi^{-i}\in \Delta^{-i}$ satisfies:
\begin{equation}
    u^i(a^{i*},\pi^{-i})\le u^i(\pi^{i*},\pi^{-i}), \forall\pi^{-i}\in \Delta^{-i},
    \label{eq:p203}
\end{equation}
then the pure strategy $a^{i*}$ is a dominated strategy of player $i$.
Although ideally no one would adopt a dominated strategy,
dominated strategies may still occur in practical problems.

\subsection{Nash Equilibrium and Bayesian Perfect Equilibrium}\label{subsec:perfect-bayesian-equilibrium}
Bayesian Perfect Equilibrium (BPE) is a sophisticated concept in game theory for analyzing dynamic games with incomplete information,
refining the Nash Equilibrium (NE) by incorporating \textit{strategies} and \textit{beliefs}.
\begin{itemize}
    \item Strategy: A plan dictating a player's actions at a given information set, based on the game's history.
    \item Belief: A probability distribution over nodes within an information set.
\end{itemize}

In real world problems, Allis~\cite{allis1994searching} gave three different definitions for solving a game.
\begin{itemize}
    \item A game is said to be ultra-weakly solved if, for the initial position(s),
    the game-theoretic value has been determined;

    \item weakly solved if, for the initial position(s),
    a strategy has been determined to obtain at least the game-theoretic value,
    for both players,
    under reasonable resources;

    \item and strongly solved if, for all legal positions,
    a strategy has been determined to obtain the game-theoretic value of the position,
    for both players,
    under reasonable resources.
\end{itemize}

In the context of Allis' definitions:
Weakly Solved equates to discovering a Nash Equilibrium strategy.
Strongly Solved corresponds to finding a Bayesian Perfect Equilibrium.

In incomplete information games,
the interdependence of strategy and belief underlines that a strategy forms an equilibrium only if the belief is mutually acknowledged.
However,
in two-player zero-sum games,
any Nash Equilibrium strategy forms an equilibrium irrespective of beliefs,
often leading to the ignore of beliefs in many algorithmic solutions.

On the information set $u$ of BPE,
the player not only has the strategy $\pi(u)$,
but also the belief $\mathfrak B(u)$ at this node.
$\mathfrak B(u)$ is a probability distribution defined on $\mathcal I(u)$.
In a two-player zero-sum game, the BPE can be strongly solved.

\subsection{Generalised Weakened Fictitious Play}\label{subsec:generalised-weakened-fictitious-play}
GWFP has similar convergence guarantees as common FP,
but allows for approximate BR strategy and perturbed average strategy updates~\cite{leslie2006generalised}.

Define the $\epsilon$-BR strategy set of Player $i$ to opponent is,
\begin{equation}
    \label{eq:p204}
    \begin{split}
        BR_{\epsilon}^{i}&\left(\Phi^{-i}\right)=\\
        &\left\{\Phi^{i} \in \Delta^{i}: R^{i}\left(\Phi^{i}, \Phi^{-i}\right) \ge R^{i}\left(BR^{i}\left(\Phi^{-i}\right), \Phi^{-i}\right)-\epsilon\right\}
    \end{split}
\end{equation}
A GWFP process is any process $\{\Phi_t\}_{n \ge 0}$ that
\begin{equation}
    \label{eq:p205}
    \Phi_{t+1} \in \left(1-\alpha_{t+1}\right) \Phi_{t}+\alpha_{t+1}\left(BR{\epsilon_{t}}\left(\Phi_{t}\right)+M_{t+1}\right)
\end{equation}

Where $\alpha_t\in[0,1]$ is the stepsize of each round of the FP,
$\{M_t\}_{t\ge 1}$ is a sequence of perturbations.
When the following conditions are met,
GWFP has the same convergence property as FP .

\begin{enumerate}
    \item $\alpha_t \rightarrow 0$ and $\epsilon_t \rightarrow 0$ as $t\rightarrow \infty$
    \item $\sum_{t\ge 1} \alpha_t=\infty $
    \item $\{M_t\}_{t\ge 1}$ satisfies for any $Q>0$:
\end{enumerate}
\begin{equation}
    \label{eq:p206}
    \lim_{t \rightarrow \infty} \sup _{k}\left\{\left\|\sum_{i=t}^{k-1} \alpha_{i+1} M_{i+1}\right\|: \sum_{i=t}^{k-1} \alpha_{i+1} \le Q\right\}=0
\end{equation}

Clearly,
a classical FP process is a GWFP process with $\epsilon_t=M_t=0$ and $\alpha_t=1/t$ for all $t$.
GWFP is a special case of online convex optimization.
showing that the convergence rate of FP is $\mathcal{O}\left(t^{-\frac{1}{2}}\right)$~\cite{abernethy2021fast}.
In addition,
GWFP can use a variety of stepsizes.
This article mainly uses the stepsizes of XFP~\cite{heinrich2015fictitious} and EFFP~\cite{hendon1996fictitious}.
The details of these stepsize are more complicated, please refer to the original paper for details.

\subsection{Bayesian Action Decoder}\label{subsec:bayesian-action-decoder}
Starting from the dynamics of human interaction,
Foerster et al.~\cite{foerster2019bayesian} highlight the cooperative nature of human players who,
by observing others' strategies,
deduce the reasons behind their actions and their implications for the environment.
This observation extends to understanding that one's actions will be interpreted by others,
enabling individuals to embed both the intended action and additional communicative signals in their behavior to enhance coordination and mutual understanding.

Building on this insight,
the BAD framework introduces a novel Markov decision process known as the public belief MDP.
This model expands upon the concept of public belief articulated by Nayyar et al.~\cite{nayyar2013decentralized},
adapting it to accommodate larger state spaces.
It categorizes the observable features into public components,
$s^{p u b}$,
such as the communal cards and historical actions in Texas Hold'em, and private components,
$s^{\text {pri }}$, like the initial hand dealt to a player.
The belief $\mathfrak{B}_t(u)$ is then refined to $\mathfrak{B}_t\left(u, s^{\text {pri }}\right)=P\left(s_t^{\text {pri }} \mid \mathfrak{U}^{p u b}\left(\sigma_u\right)\right)$,
where $\mathfrak{U}^{p u b}\left(\sigma_u\right)$ encapsulates all publicly available information along a given trajectory.
Throughout this iterative process,
an agent $i$ estimates its expected reward as:
\begin{equation}
    \label{eq:p207}
    R_{u_i}(\pi)=\sum_{s^{pri}_i} \mathfrak B_t\left(u_i,s^{pri}_i\right)R\left(s^{pri}_i,\pi\right)
\end{equation}

  \section{Belief Update Fictitious Play}\label{sec:method}

\subsection{Belief Update in Extensive From Game}\label{subsec:BUFP}
The core of BUFP is to embed belief update into the iterations of extensive-form game.
We define $\sigma_{u}=\left((u_{1}, a_{1}),(u_{2}, a_{2}),\dots ,(u_{-1}, a_{-1}) \right), u \in \mathcal{U}$ as the information set-action pair sequence,
in which $\sigma_u[j]$ represents the $j$-th information set-action pair $(u_j,a_j)$ of this sequence,
$\sigma_u[:j]$ represents the $j$ information set-action pairs before this sequence $((u_1,a_1),\dots,(u_{j-1},a_{j-1})) $.
Let $\sigma_u+a$ denote the sequence of extending $\sigma_u$ with information set-action $(u,a)$,
$\mathfrak{U}(\sigma_u)=\{u_1,u_2,\dots\}$ represents the set of information sets of all experiences of the sequence of $\sigma_u$.

Definition a realization plan of player $i \in \mathcal{N}$ is a function,
$x:\sigma ^i\rightarrow [0,1]$,
such that $x(\emptyset)=1$ and $\forall \sigma_{u \in \mathcal{U}^i}: x(\sigma_u) = \sum_{a\in \mathcal{A}(u)} x(\sigma_u+a)$.
Behavioral strategy $\pi$ can be transformed into implementation plan $x_{\pi}\left(\sigma_{u}\right)=\prod_{\left(u^{\prime}, a\right) \in \sigma_ {u}} \pi\left(u^{\prime}, a\right)$,
where the symbol $\pi(u', a)$ represents the probability of taking action $a$ on the information set $u^\prime$.
Similarly,
the realization plan $x$ can also be converted into a behavioral strategy, $\pi(u,a)=\frac{x(\sigma_u+a)}{x(\sigma_u)}$, if $x(\sigma_u) =0$ then $\pi$ is randomly defined.

For an information set-action sequence $\sigma_u$,
the subsequent counterfactual subgame is $L(\sigma_u)$.
The form of the counterfactual subgame tree has not changed,
only the players’ beliefs have changed in this counterfactual subgame tree.
Define the fictitious behavior strategy $\pi_{L(\sigma_u)}$:

\begin{equation}
    \label{eq:p401}
    \pi_{L(\sigma_u)}(u^\prime,a)=
    \begin{cases}
        \pi(u^\prime,a) & (u^\prime,a) \notin \sigma_u \\
        \pi(u^\prime,a) & (u^\prime,a) \in \sigma_u \text{ and } P(u^\prime)\neq P(u)\\
        1 & \text{otherwise}
    \end{cases}
\end{equation}

Like CFR,
only player $P(u)$ arrives at this information set intentionally,
and the rest of the players follow the normal strategy.
Define the BR strategy on the counterfactual subgame $L(\sigma_u)$ as:

\begin{equation}
    \label{eq:p402}
    b^{*i}(\pi^{-i}_{L(\sigma_u)})=\mathop{\arg\max}\limits_{\gamma\in\mathcal{A}(u)}R_{u}^i(\pi_{L(\sigma_u)}^i|_{u\rightarrow \gamma},\pi^{-i}_{L(\sigma_u)})
\end{equation}

Where $P(u)=i$, $\pi|_{u\rightarrow \gamma}$ represents that the information set strategy remains unchanged except that the action $\gamma$ is taken on the information set $u$.
$R_{\pi^{-i},u}$ represents that in the information set $u$ and subsequent sequences,
when all players except player $i$ take actions according to the behavioral strategy $\pi$,
player $i$ the optimal response.

Similar to BAD, our payoff is also calculated based on belief:
\begin{equation}
    \label{eq:p403}
    \begin{aligned}
        R_{u}^i&(\pi_{L(\sigma_u)}^i|_{u\rightarrow \gamma},\pi^{-i}_{L(\sigma_u)})= \\
        &\sum_{s^{pri}_i\in \mathcal I(u)}\mathfrak{B}_t(u_i,s^{pri}_i) R_{s^{pri}}^i(\pi_{L(\sigma_u)}^i|_{u\rightarrow \gamma},\pi^{-i}_{L(\sigma_u)})
    \end{aligned}
\end{equation}
Update strategy:
\begin{equation}
    \label{eq:p404}
    \pi_{t+1}(u)=(1-\alpha_t)\pi_{t}(u)+\alpha b^{*i}(\pi^{-i}_{L(\sigma_u)})
\end{equation}
Then update the belief:
\begin{equation}
    \label{eq:p405}
    \mathfrak{B}_{t+1}(u_i,s^{pri}_i)=\frac{x_{\pi_{t+1}}(\sigma_{s_i^{pri}})}{\sum_{s_i^{\prime pri}}x_{\pi_{t+1}}(\sigma_{s_i^{\prime  pri}})}
\end{equation}
When we adopt the iteration stepsize of EFFP,
as the iterations continue,
the average strategy $\bar{\pi}_{t+1}$ and belief $\mathfrak{B}$ will together form a BPE.

\subsection{BUFP is Equivalent to GWFP}\label{subsec:123}

In this section,
we utilize the stepsize of EFFP as an example to demonstrate that BUFP qualifies as a variant of GWFP.
For the GWFP process to hold, i
t must satisfy $\alpha_t \rightarrow 0$ and $\epsilon_t \rightarrow 0$ as $n\rightarrow \infty$,
ensure $\sum_{t\ge 1} \alpha_t=\infty$,
and for any $Q>0$,
adhere to formula~\ref{eq:p205}.
We will sequentially verify that BUFP(EF) meets these prerequisites.
Given that GWFP is applicable solely to normal-form games,
an initial step involves examining the shifts in mixed equilibrium of normal-form games attributable to the BUFP process.

\begin{equation}
    \label{eq:p406}
    \pi_{t+1}(u)=\frac{t}{t+1}\pi_{t}(u)+\frac{1}{t+1} b^{*i}(\pi^{-i}_{L(\sigma_u)})
\end{equation}

Here, $b^{*i}(\pi^{-i}_{L(\sigma_u)},u)$ is succinctly denoted as $b_t^*(u)$.
Assuming the terminal state $z$ is reached within a maximum of $l$ steps, i.e., $\max_{z\in \mathcal{Z}} |\sigma_z|=l$,
we have:
\begin{equation}
    \label{eq:p407}
    x_{\pi_{t+1}}(\sigma_z)=\prod_{(u,a)\in \sigma_u}\pi_{t+1}(u,a)
\end{equation}

Incorporating $\pi_{t+1}(z)$ yields:
\begin{equation}
    \label{eq:p408}
    x_{\pi_{t+1}}(\sigma_z)=\prod_{(u,a)\in\{\sigma_u\}} \left( \frac{t}{t+1}\pi_{t}(u,a)+ \frac{1}{t+1}b^*_t(u,a) \right)
\end{equation}
where $x{\theta_t}$ encapsulates the subsidiary terms of both $\pi_t(u,a)$ and $b^t(u,a)$.
\begin{equation}
    \label{eq:p409}
    \begin{split}
        x_{\pi_{t+1}}(\sigma_z)=&\left( \frac{t}{t+1} \right)^l x_{\pi_t}(\sigma_z)+\\
        &\left(\frac{1}{t+1}\right)^l x_{b_t^*}(\sigma_z)+\\
        &\frac{(t+1)^l-t^l-1}{(t+1)^l} x_{\theta_t}(\sigma_z)
    \end{split}
\end{equation}
which implies:
\begin{equation}
    \label{eq:p410}
    \Pi_{t+1}=\left( \frac{t}{t+1} \right)^l \Pi_t+\left(\frac{1}{t+1}\right)^l B_t^*+\frac{(t+1)^l-t^l-1}{(t+1)^l}\Theta_t
\end{equation}

From which we deduce:
\begin{equation}
    \label{eq:p411}
    \alpha_t=\frac{t^l-(t-1)^l}{t^l}=1-\left( \frac{t-1}{t}\right)^l\\
\end{equation}

\begin{equation}
    \label{eq:p412}
    M_t=\frac{(\Theta_t-B^*_t)((t+1)^l-t^l-1)}{(t+1)^l-t^l}
\end{equation}

\begin{equation}
    \label{eq:p413}
    \begin{aligned}
        \Pi_{t+1}&=\left( \frac{t}{t+1} \right)^l \Pi_t+\left(\frac{1}{t+1}\right)^l B_t^*+\frac{(t+1)^l-t^l-1}{(t+1)^l}\Theta_t\\
        &=(1-\alpha_{t+1}) \Pi_t+\alpha_{t+1}\left( B^*_t+ M_t\right)
    \end{aligned}
\end{equation}
Thus, we establish that BUFP(EF) aligns with the GWFP framework, fulfilling the same convergence criteria as FP.

\subsubsection{Proof of Condition 1}
Regarding the condition for $\alpha_t\rightarrow 0$,
given that the game unfolds across $l$ levels,
analysis of the preceding equation reveals that the update step $\alpha_t$ is formulated as:
\begin{equation}
    \alpha_t=\frac{t^l-(t-1)^l}{t^l}=1-\left( \frac{t-1}{t}\right)^l
    \label{eq:alpha}
\end{equation}
which naturally fulfills the condition $\lim_{t\rightarrow\infty} \alpha_t = 0$,
thereby demonstrating convergence as $t$ approaches infinity.

\subsubsection{Proof of Condition 2}
For the proof of Condition 2, we evaluate the sum of $\alpha_{t}$ as follows:
\begin{equation}
    \begin{aligned}
        \sum_{t \geq 1} \alpha_{t} &=T-\sum_{t=1}^{T}\left(\frac{t-1}{t}\right)^{l} \\
        & \geq T-\sum_{t=1}^{T}\left(\frac{t-1}{t}\right)\left(\frac{t}{t+1}\right) \cdots\left(\frac{t+l-2}{t+l-1}\right) \\
        &=T-\sum_{t=1}^{T}\left(\frac{t-1}{t+l-1}\right) \\
        &=l \sum_{t=1}^{T}\left(\frac{1}{t+l-1}\right)
    \end{aligned}
    \label{eq:p414}
\end{equation}
Given the harmonic series $\sum_{t}^{\infty} 1/t = \infty$,
it naturally follows that for any finite $k$, $k \sum_{t=1}^{T}\left(\frac{1}{t+l-1}\right)$ diverges to infinity, thereby satisfying Condition 2.

\subsubsection{Proof of Condition 3}
Given the equation:
\begin{equation}
    \label{eq:p415}
    M_{t+1}=\frac{(\Theta_t-B^*_t)((t+1)^l-t^l-1)}{(t+1)^l-t^l}
\end{equation}
we can then express equation$~$\ref{eq:p205} as follows:
\begin{equation}
    \label{eq:p416}
    \begin{aligned}
        &\left\| \sum_{i=t}^{k-1} \alpha_{i+1}M_{i+1} \right\| \\
        =&\left\| \sum_{i=t}^{k-1} \frac{(i+1)^l-i^l}{(i+1)^l}\frac{(\Theta_i-B^*_i)((i+1)^l-i^l-1)}{(i+1)^l-i^l} \right\| \\
        =&\left\|\sum_{i=t}^{k-1}(\Theta_i-B^*_i)\left(1-\frac{i^l+1}{(i+1)^l}\right)\right\|\\
    \end{aligned}
\end{equation}

As $t \to \infty$,
the term $\left(1 - \frac{i^l + 1}{(i+1)^l}\right)$ approaches 0.
Given $\sum{i=t}^{k-1} \alpha_{i+1} \leq Q$ implies $k$ is finite,
the expression approaches 0, implying that $\alpha_{i+1}M_{i+1}$ converges to 0.
Therefore,
condition 3 is naturally satisfied, indicating that BUFP(EF) aligns with the characteristics of GWFP and shares the same convergence properties as FP.

  \section{Experiments and Analysis}\label{sec:experiments-and-analysis}

\begin{figure*}[t]
    \centering
    \includegraphics[width=\linewidth]{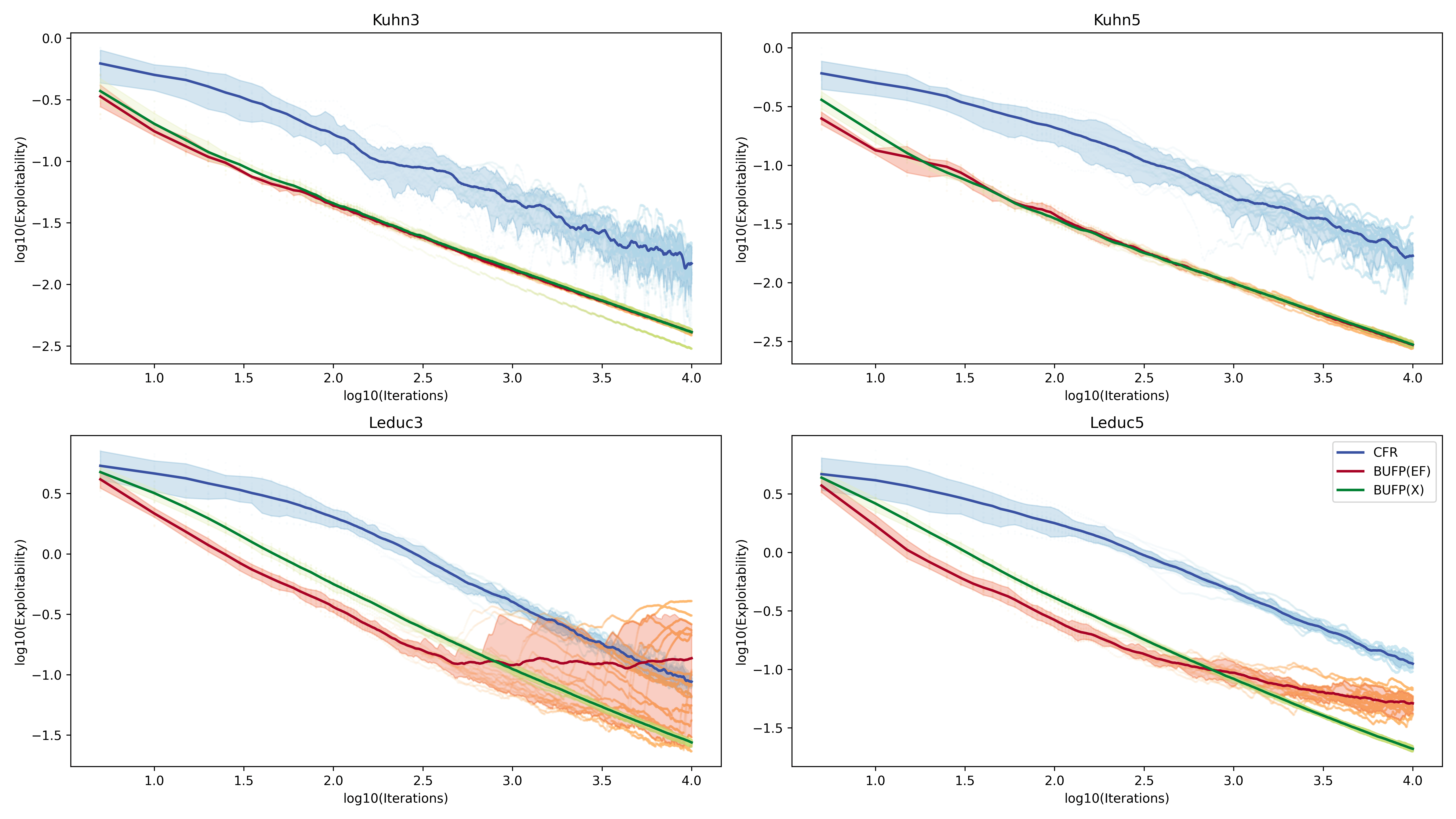}
    \caption{
        In the depicted figure,
        the horizontal axis denotes the number of iterations,
        while the vertical axis quantifies the strategy's \textbf{exploitability}, i.e., its deviation from Nash Equilibrium.
        For each algorithm under consideration,
        30 experiments were conducted.
        The shaded area represents the 90\% confidence interval for each algorithm's performance.
    }
    \label{fig:nc}
\end{figure*}

\begin{figure*}[t]
    \centering
    \includegraphics[width=\linewidth]{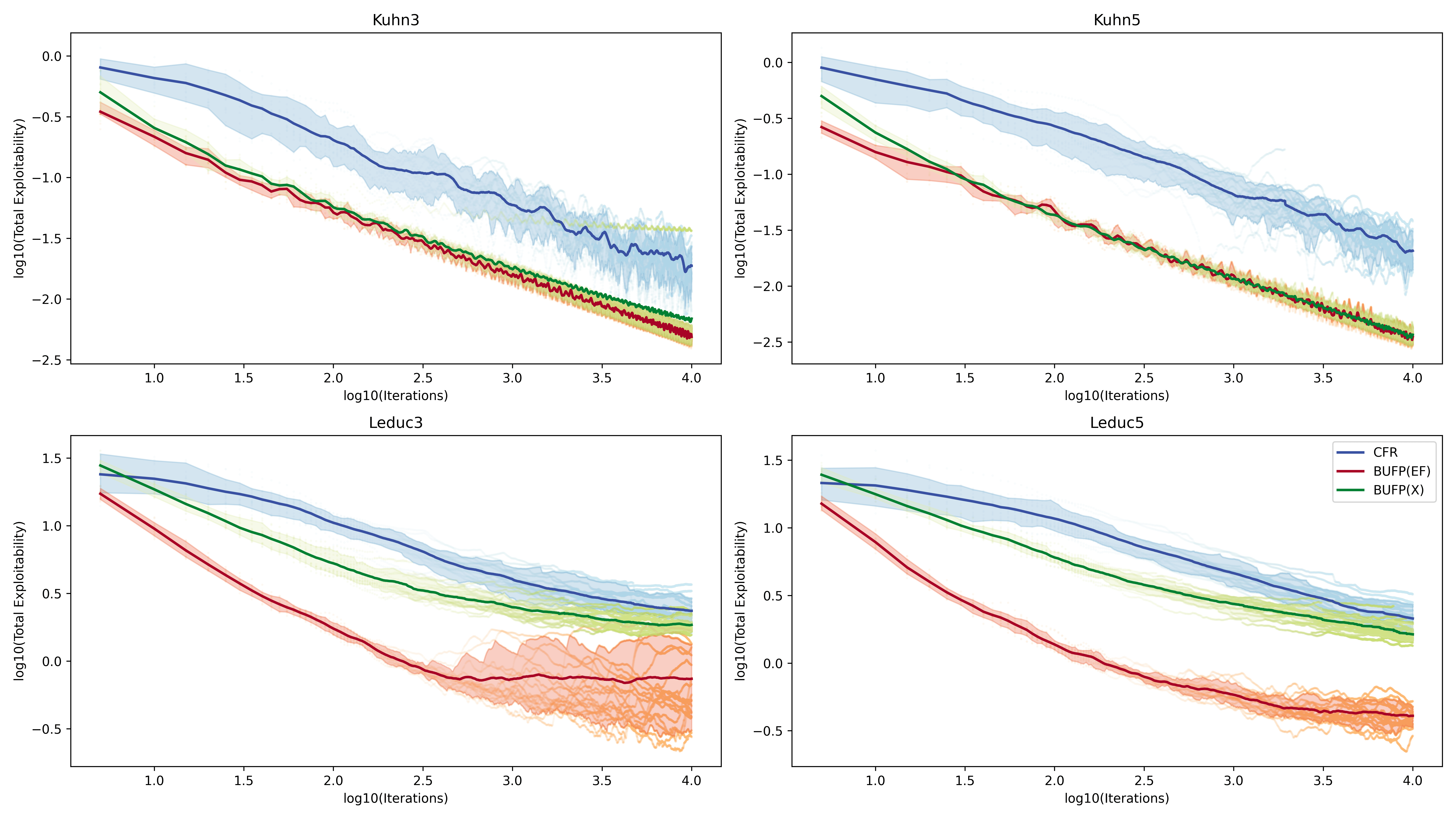}
    \caption{
        In the depicted figure,
        the horizontal axis denotes the number of iterations,
        while the vertical axis quantifies the strategy's \textbf{total exploitability}, i.e., its deviation from Bayesian Perfect Equilibrium.
        For each algorithm under consideration,
        30 experiments were conducted.
        The shaded area represents the 90\% confidence interval for each algorithm's performance.
    }
    \label{fig:pc}
\end{figure*}

\subsection{Algorithm Convergence Rate}\label{subsec:234}
In our study,
we concentrate on Kuhn poker~\cite{kuhn1950simplified} and Leduc poker~\cite{southey2012bayes},
both widely recognized benchmarks in game theory research.
Our simulations adhere to the rlcard framework,
leveraging exploitability and total exploitability metrics to assess algorithmic convergence.
Here, exploitability quantifies the deviation of a strategy from NE,
while total exploitability measures its divergence from PBE,
following the methodology described in previous research~\cite{brown2020equilibrium}.

Figure ~\ref{fig:nc} illustrates that BUFP(X) achieves a notably faster convergence rate than both CFR and BUFP(EF),
likely due to BUFP's comprehensive assessment of opponent hand beliefs during the iterative process, unlike CFR's incremental approach to hand distribution approximation.

Interestingly,
despite BUFP(EF)'s consistent progress towards PBE,
it experiences considerable variability when evaluated against the NE standard,
indicating that BUFP(EF) may not excel in scenarios where NE serves as the primary benchmark for convergence.

\subsection{Bayesian Perfect Equilibrium}\label{subsec:bayesian-perfect-equilibrium}
\begin{table*}[h]
    \centering
    \caption{Winnings in mb/h for initial node}
    \label{tab:NE}
    \begin{tabular}{|r||r|r|r|r|r|}
        \hline                   & CFR P2 & BUFP(EF) P2 & BUFP(X) P2 & P1 Average \\
        \hline \hline CFR P1     & 19.21  & 20.17       & 17.93      & 19.10      \\
        \hline BUFP(EF) P1       & 18.34  & 19.02       & 17.71      & 18.36      \\
        \hline BUFP(X) P1        & 20.38  & 20.68       & 19.10      & 20.06      \\
        \hline \hline P2 Average & 19.31  & 19.96       & 18.25      &            \\
        \hline
    \end{tabular}
\end{table*}

\begin{table*}[h]
    \centering
    \caption{Winnings in mb/h for non-equilibrium(dominated) path}
    \label{tab:non_NEP}
    \begin{tabular}{|r||r|r|r|r|r|}
        \hline                   & CFR P2 & BUFP(EF) P2 & BUFP(X) P2 & P1 Average \\
        \hline \hline CFR P1     & -33.86 & -44.54      & -22.02     & -33.47     \\
        \hline BUFP(EF) P1       & -5.15  & -20.03      & 4.49       & -6.89      \\
        \hline BUFP(X) P1        & -4.89  & -22.37      & 3.94       & -7.75      \\
        \hline \hline P2 Average & -14.63 & -28.98      & -4.52      &            \\
        \hline
    \end{tabular}
\end{table*}

Any AI(human) is remains the likelihood of deviating into non-equilibrium (dominated) paths due to errors in judgment.
In such situations,
the aim is for AI to offer strategic guidance.
While neither CFR nor BUFP(X) ensure optimal decisions on these dominated paths, the importance of BUFP(EP) becomes evident.
Figure~\ref{fig:pc} shows that although BUFP(EP) may converge more slowly initially,
it attains the fastest rate in minimizing exploitability across the full information set.

Using 5-Leduc poker as an example,
we highlight the advantages of BUFP(EP) in dealing with non-equilibrium paths.
After training CFR, BUFP(EF), and BUFP(X) for 100,000 iterations,
their exploitabilities were measured at $5.15$, $5.33$, and $3.16$ (mb/h) respectively,
indicating similar performances against optimal strategies but differing significantly in handling dominated strategy paths.
By pitting these three AIs against one another,
we calculate the interplay of payoffs.
Tables ~\ref{tab:NE} and ~\ref{tab:non_NEP} display the payoffs for player 1 under varying initial conditions:
Table ~\ref{tab:NE} for the initial node and Table ~\ref{tab:non_NEP} for scenarios where player 1 is forced to play Call on the first decision state.
It’s no doubt that being forced into a strategy in poker is always a dominated strategy.

From the Table~\ref{tab:NE} and Table~\ref{tab:non_NEP},
we can see that in the original game,
if Player 1 adopts the CFR strategy,
the average profit obtained is 19.10.
When the game moves into an dominated strategy path,
and Player 1 continues to adopt the CFR strategy,
Player 1's profit decreases by 53.57, reaching -33.47.
At the same time,
in the original game,
if Player 1 adopts the BUFP(EF) strategy,
the average profit obtained is 18.36.
When the game moves into an dominated strategy path,
and Player 1 continues to adopt the BUFP(EF) strategy,
Player 1's profit decreases by 25.25, reaching -6.89.
This indicates that if a previous player (agent) misjudges and enters an dominated strategy path,
the decline in profit from BUFP(EP) is the least,
and it can recover the most losses when a previous action is mistaken,
with the recovered loss exceeding that of CFR by $(53.57-25.25)/53.57=52.87\%$.

Additionally,
in the original game,
if Player 2 adopts the CFR strategy,
the average profit obtained is 19.31.
When the game moves into an dominated strategy path,
and Player 2 continues to adopt the CFR strategy,
Player 2's profit increases by 33.94,
reaching -14.63 (since it is a zero-sum game, the lesser the profit for Player 2, the better).
At the same time, in the original game,
if Player 2 adopts the BUFP(EF) strategy,
the average profit obtained is 19.96.
When the game moves into an dominated strategy path,
and Player 2 continues to adopt the BUFP(EF) strategy,
Player 2's profit increases by 48.94,
reaching -28.98.
This indicates that if the opponent player (agent) misjudges and enters an dominated strategy path,
the increase in profit from BUFP(EP) is the least, and it can secure the most excess profit, thereby maximizing the punishment of the opponent's mistake,
with the secured profit exceeding that of CFR by $(48.94-33.94)/33.94=44.20\%$.
In summary, when the opponent (oneself) enters an dominated strategy path,
BUFP(FP) can on average secure $48.53\%$ more excess profit (recover $48.53\%$ more losses).

  \section{Conclusion}\label{sec:conclusion}

In our research,
we introduce an innovative version of FP designed specifically for extensive-form games,
which we call Belief Update Fictitious Play (BUFP).
BUFP uniquely integrates the belief update mechanism from the BAD framework,
significantly enhancing FP's versatility for various game types.
Through meticulous adjustment of the training stepsize,
BUFP(EF) methodically advances towards BPE,
ensuring consistent convergence to equilibrium and achieving optimal outcomes across diverse states in two-player zero-sum games.
Concurrently, BUFP(X) achieves a convergence rate on par with CFR, highlighting its efficiency.

Our future endeavors will focus on enhancing the BUFP methodology by incorporating a wider spectrum of FP variants,
including Monte Carlo (MC) and Follow the Regularized Leader (FTRL) approaches.
This expansion aims to significantly broaden the algorithm's versatility and scope of application.
Moreover, we acknowledge the efficiency challenges faced by BUFP(EF) in converging to PBE.
Addressing this,
we will also dedicate efforts towards optimizing the efficiency of BUFP(EF),
seeking innovative solutions to improve its performance in solving PBE more effectively and swiftly.
Through these initiatives,
we aim to refine BUFP's operational efficiency while extending its utility across a more diverse set of game-theoretic challenges.


\end{document}